\documentclass[aps,pra,reprint,superscriptaddress]{revtex4-1}

\usepackage{graphicx}
\usepackage{dcolumn}
\usepackage{bm}
\usepackage{hyperref}

\begin{document}


\title{Current Induced Nuclear Spin Depolarization at Landau Level Filling Factor $\nu$=1/2}


\author{Y.~Q. Li}
\email[]{Email address: yqli@iphy.ac.cn}
\affiliation{Max-Planck-Institute for Solid State Research, Heisenbergstra\ss e 1, D-70569, Stuttgart, Germany}
\affiliation{Institute of Physics, and Beijing National Laboratory for Condensed Matter Physics, Chinese Academy of Sciences, Beijing 100190, China}


\author{V. Umansky}
\affiliation{Weizmann Institute of Science, Rehovot 76100, Israel}

\author{K. von Klitzing}
\author{J.~H. Smet}
\email[]{Email address: j.smet@fkf.mpg.de}
\affiliation{Max-Planck-Institute for Solid State Research, Heisenbergstra\ss e 1, D-70569, Stuttgart, Germany}

\date{\today}

\begin{abstract}
Hyperfine interactions between electron and nuclear spins in the quantum Hall regime provide powerful means for manipulation and detection of nuclear spins. In this work we demonstrate that significant changes in nuclear spin polarization can be created by applying an electric current in a 2-dimensional electron system at Landau level filling factor $\nu=1/2$. Electron spin transitions at $\nu=2/3$ and $1/2$ are utilized for the measurement of the nuclear spin polarization.  Consistent results are obtained from these two different methods of nuclear magnetometry. The finite thickness of the electron wavefunction is found to be important even for a narrow quantum well. The current induced effect on nuclear spins can be attributed to electron heating and the efficient coupling between the nuclear and electron spin systems at $\nu$=1/2. The electron temperature, elevated by the current, can be measured with a thermometer based on the measurement of the nuclear spin relaxation rate. The nuclear spin polarization follows a Curie law dependence on the electron temperature.  This work also allows us to evaluate  the electron $g$-factor in high magnetic fields as well as the polarization mass of composite fermions.

\end{abstract}

\pacs{73.40.-c, 73.20.-r, 73.63.Hs}


\maketitle


\section{\label{sec:intro}Introduction}

In GaAs and many other semiconductors, electron spins and nuclear spins interact with each other via the hyperfine interaction~\cite{Dyakonov84}. This interaction forms the basis for many ingenious methods to detect and manipulate nuclear spins via the electronic states~\cite{Li08,Kalevich08,Reilly08,ZhangYJ11}.
Conversely, nuclear spins can also be used to probe and study electronic states in molecules and various condensed matter systems. For instance, the electron spin polarization can be measured with the Knight shift in nuclear magnetic resonance (NMR) experiments~\cite{Barrett95,Stern04,Dementyev99,Melinte00,Knumada07,Tiemann12}. Nuclear spin relaxation measurements yield important information about the electronic systems~\cite{Berg90,Smet02,Hashimoto02,Spielman05,Kumada05,Tracy06,Kumada06,Tracy07,Zhang07,Li09}.

An essential ingredient of a nuclear spin relaxation measurement is to first drive the nuclear spin system out of equilibrium.
NMR provides the most direct way for eliciting changes in the degree of nuclear spin polarization. The RF radiation in NMR experiments, however, unavoidably raises the electron temperature in the sample, in particular when operating at dilution refrigerator temperatures. This is detrimental to the fragile states such as those formed as a result of electron correlations in the fractional quantum Hall regime. Fortunately, alternative methods for manipulating nuclear spin polarization are available. They are based on the spin flip-flop term of the hyperfine interaction:
\begin{equation}
\label{eq:flip-flop}
H_\mathrm{flip-flop}=\frac{1}{2}A_\mathrm{HF}({\hat{I}_{+}}\cdot \hat{S}_{-}+{\hat{I}_{-}}\cdot \hat{S}_{+}),
\end{equation}
where $A_\mathrm{HF}$ is the hyperfine constant, and  $\hat{I}_+$($\hat{I}_-$) and $\hat{S}_+$($\hat{S}_-$) are raising (lowering) operators for the nuclear and electron spins, respectively. This term describes processes in which the flip of an electron spin simultaneously triggers the reversal of a nuclear spin. Driving the electron spin system out of thermal equilibrium by an external source (e.g.\ microwave, light, or electric current) will cause the electron spins to relax back. The electron relaxation is accompanied by polarization of the nuclear spins. This dynamic nuclear polarization (DNP) process has been realized in numerous experiments, including optical pumping by circularly polarized light, \cite{Barrett95} electron spin resonance (ESR)~\cite{Dobers88}, inter-edge channel scattering in the quantum Hall regime~\cite{Dixon97}, current induced scattering near  Landau level filling factor $\nu=$2/3~\cite{Kronmueller98,Kronmueller99,Smet01}, and other fractional fillings~\cite{Kraus02,Stern04,Kou10}. It also occurs in the breakdown regime of the integer and fractional quantum Hall effects~\cite{Kawamura07,Dean09,Kawamura09}.

In this work we highlight a different method for manipulating the nuclear spin polarization. It does not rely on the aforementioned conventional dynamic nuclear polarization processes. It will be referred to as electrically controlled thermal depolarization. It is based on current induced heating in the two-dimensional electron system (2DES) when a partially polarized composite fermion liquid at half filling of the lowest Landau level forms. The strong hyperfine interaction transfers energy from electrons to nuclear spins and hence raises the entropy as well as the temperature of the nuclear spin system. Significant changes in nuclear spin polarization can be obtained with low current densities. In contrast to previously reported techniques for electrical control of the nuclear spin polarization, this $\nu=1/2$ based technique can in principle produce spatially homogeneous changes in the nuclear spin polarization across a large area. The changes in nuclear spin polarization are measured with two different methods. One is based on the spin transition in the $\nu=1/2$ state itself~\cite{Tracy07,Li09}, and the other relies on the spin phase transition at $\nu=2/3$~\cite{Kronmueller98,Smet04}.  The nuclear spin depolarization induced by the current can be described by a Curie law.

This paper is organized as follows. In Sec.\,\ref{sec:theory}, we give a brief description of the composite fermion picture and the spin transitions at fillings 1/2 and 2/3. Sec.\,\ref{sec:Methods} is devoted to the experimental details, including the sample preparation, the measurement setup, and the principles of the nuclear magnetometry using the 1/2 and 2/3 spin transitions. Effects associated with the finite thickness of the two-dimensional electron system will also be discussed in this section.  In Sec.\,\ref{sec:Result} the main experimental results will be presented. They include the current induced effects on the nuclear spins and electron transport properties. The mechanism for electrical controlled nuclear spin depolarization will be discussed based on measurements of electron temperatures. Finally, concluding remarks will be given in Sec.\,\ref{sec:summary}.

\section{\label{sec:theory}Theoretical background}

For a 2D electron system with density $n_s$ subject to a perpendicular magnetic field $B$,  all electrons reside in the lowest Landau level when $B$ exceeds $n_s h/e$ (i.e.\, Landau level filling factor $\nu\equiv\frac{n_s/h}{eB}<1$), where $n_s$ is the density of the 2DES, $h$ is the Planck constant, and $e$ is the electron's charge. The orbital degree of freedom is no longer relevant and the physics is governed by electron-electron interactions. They give rise to a large number of fractional quantum Hall states when the disorder is sufficiently weak~\cite{DasSarma97}. The many-body wavefunctions proposed by Laughlin provide a solid foundation for understanding the nature of these states~\cite{Laughlin83}. It is however also possible to describe the appearance of these correlated fractional quantum Hall ground states in an intuitive, single particle picture by introducing quasi-particles referred to as composite fermions~\cite{Jain07,HLR93}.

A composite fermion (CF) comprises one electron and an even number ($q=2,4$) of magnetic flux quanta. CFs no longer experience the external magnetic field, but a drastically reduced effective magnetic field which in a mean field approximation~\cite{HLR93} is given by $B_\mathrm{eff}=B-q n_s h/e$ and vanishes exactly at even denominator filling $1/q$. At this filling composite fermions form a compressible Fermi sea which in many ways resembles the 2D electron Fermi liquid at zero magnetic field. At filling factors away from 1/q, the Landau quantization of composite fermions in a non-zero $B_\mathrm{eff}$ gives rise to the integer quantum Hall effect of composite fermions. It is equivalent to the fractional quantum Hall effect at fillings $p/(pq\pm 1)$, where $p$ is the number of filled CF Landau levels. For example, the $\nu=2/3$ state corresponds to the integer quantum Hall state of composite fermions with two attached flux quanta when two CF Landau levels are completely filled.

\subsection{\label{subsec:SpinTransition}Spin transitions at $\nu=2/3$ and $1/2$}

\begin{figure}
\includegraphics*[width=7.5 cm]{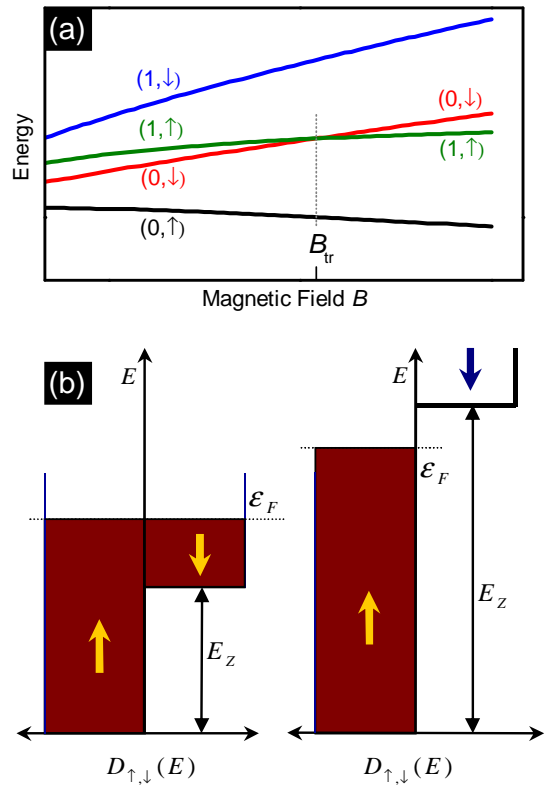}
\caption{\label{Fig1_SpinTransition}(a) Energy level diagram for filling factor 2/3. The spin transition field $B_\mathrm{tr}$ is marked with the dotted line. When $B<B_\mathrm{tr}$, the ground state is spin unpolarized with the composite fermion Landau levels (0,$\uparrow$) and (0,$\downarrow$) occupied, while for $B>B_\mathrm{tr}$ the two filled levels are (0,$\uparrow$) and (1,$\uparrow$), and the ground state becomes fully polarized; (b) Spin transition for a non-interacting and disorder free composite fermion system at $\nu$=1/2. The density of states solely depends on the composite fermion effective mass. Transition from partial to full spin polarization takes place when the Zeeman energy surpasses the Fermi energy.}
\end{figure}

The cyclotron mass and Landau quantization energy of composite fermions are not related to the conduction band mass of  the electrons~\cite{Park98}. This has some important implications for the spin related physics in the fractional quantum Hall regime. In the following, we only discuss two cases which will be important for the nuclear magnetometry carried out in this work. For simplicity, we start our discussion  with an ideal 2D electron system, whose wavefunction has zero spread in the growth direction. We also do not consider the effect of nuclear spin polarization. It will be treated in a subsequent subsection.

At filling factor 2/3, composite fermions experience an effective (perpendicular) magnetic field of $-B/3$. The energy spectrum is quantized into a ladder of CF Landau levels ($n=0,1,2,...$) with a spacing of $\hbar\omega_\mathrm{CF} = \hbar \frac{e}{m_\mathrm{CF}}\frac{B}{3}=\frac{\hbar e}{3\xi m_e}\sqrt{B}$, where $m_e$ is the free electron mass, and the composite fermion mass is written as $m_\mathrm{CF}=\xi\sqrt{B}m_e$~\cite{Park98}. Each of the CF Landau levels is split into two spin sub-levels ($s=\uparrow,\downarrow$) by the Zeeman energy $E_Z=g_e\mu_B B_\mathrm{tot}$, where $B_\mathrm{tot}$ is the total magnetic field. Note the effective field only controls the orbital degree of freedom of the composite fermions. As shown in Fig.\,1(a), energy levels (0,$\downarrow$) and (1,$\uparrow$) cross each other at perpendicular field $B=B_\mathrm{tr}$ due to the different field dependencies of $E_C$ and $E_Z$. When $B<B_\mathrm{tr}$, the two occupied levels,  (0,$\uparrow$) and (0,$\downarrow$), have opposite spin orientation, so the ground state is spin unpolarized. In contrast, when $B>B_\mathrm{tr}$ the two filled levels, (0,$\uparrow$) and(1,$\uparrow$), have identical spin orientations in Fig. 1(a) and the ground state is fully spin polarized. The spin transition field $B_\mathrm{tr}$ satisfies
\begin{equation}
\label{eq:Btr_two3rds}
B_{\mathrm{tr}}|_{\lambda=0, B_N=0}=\frac{4}{9}(\frac{1}{g_e\xi})^2 \cos^2\theta,
\end{equation}
where $\theta$ is the tilt angle, i.e.  the angle between the total field $B_\mathrm{tot}$ and the perpendicular field $B$. Here the indices $\lambda=0$ and $B_N=0$ indicate that we are dealing with a special case of zero thickness of the electron wavefunction and zero nuclear spin polarization. More general cases will be discussed in subsequent sections.

For the Fermi sea at $\nu=1/2$, the composite fermion spins are not always fully polarized as well~\cite{Kukushkin99,Dementyev99,Melinte00}. A transition from partial to full spin polarization takes place when the Zeeman energy exceeds the Fermi energy of the composite fermions. In the simplest model, the composite fermions are treated as non-interacting particles and the disorder is ignored. Under these assumptions, the Fermi energy can be written as $\varepsilon_F=\hbar^2 k_F^2/(2m_\mathrm{CF})$, where $k_F$ is the CF Fermi wavevector. It is straightforward to obtain that when the (perpendicular) magnetic field $B \geq (\frac{1}{g_e\xi} )^2 \cos^2\theta$, the composite fermion sea becomes fully spin polarized (see Fig.\,1(b)). It is interesting to note that the spin transition field at $\nu=2/3$ only differs from that at $\nu=1/2$ by a factor of $4/9$.

A useful feature of the partially polarized Fermi sea at $\nu=1/2$ is that the energy spectrum is continuous near the Fermi energy for both spin populations. A spin flip of a composite fermion may require a change in orbital momentum, but costs no or very little energy just like a nuclear spin flop. Hence, the interaction between nuclear spins and electron spins is expected to be strong. The situation resembles the Korringa type  of nuclear spin-electron spin interaction in simple metals~\cite{Korringa50,Tracy07}. Indeed, in experiments the nuclear spin relaxation time has been found to be as short as $\sim$100\,s even at temperatures below $30$\,mK~\cite{Li09}. Therefore, the $\nu=1/2$ state lends itself to manipulate nuclear spins because of the efficient coupling between these two spin systems. This will be the central theme of this paper.

\subsection{\label{subsec:FiniteThickness}Finite thickness effect}
It should be pointed out that the transition fields given in the previous subsection are calculated for truly 2D systems. For a 2DES which forms in a GaAs/AlGaAs quantum well or heterostructure, the finite thickness of the electron wavefunction in the growth direction softens the Coulomb potential. When the magnetic length, $l_B=\sqrt{\hbar/eB}$, becomes comparable to the thickness of the electron wavefunction, the ratio $\eta=E_Z/E_C$, which determines the spin transition fields for both $\nu=2/3$ and $1/2$, is no longer proportional to $\sqrt{B}$. In the strong $B$ limit, the Coulomb potential scales with $\log B$ instead of $\sqrt{B}$. As will be shown in Sec.\,\ref{subsec:Two3rds}, the finite thickness effect significantly modifies the spin transition field even for a narrow quantum well.

A precise treatment of the finite thickness effect requires numerical calculations~\cite{Davenport12}, which are beyond the scope of this paper. Here we follow an approach introduced by Zhang and Das Sarma~\cite{Zhang86}. According to their theoretical study, a modified Coulomb potential of the form  $V(r)\propto (r^2+\lambda^2)^{-1/2}$ provides a reasonably good approximation to more rigorous numerical calculations. Here $\lambda$ is a length parameter that can be viewed as the effective thickness of the electron wavefunction. In this model the Coulomb energy follows $E_C\propto(l_B^2+\lambda^2)^{-1/2}$, instead of $E_C\propto l_B^{-1}$ in the zero thickness limit.  For analyzing experiments, it is useful to define $B_\lambda=\hbar/(e\lambda^2)$, and rewrite the Coulomb energy as
\begin{equation}
E_C=\frac{1}{4\pi\epsilon\epsilon_0}\frac{e^2}{l_B}\sqrt{\frac{B_\lambda}{B+B_\lambda}}\propto \sqrt{\frac{B_\lambda B}{B+B_\lambda}} .
\end{equation}
If the tilt angle $\theta$ is small so that the orbital effect of the in-plane field is negligible, the spin transition field, taking into account the non-zero thickness of the 2DES, can be written as
\begin{equation}
\label{eq:Btr_finite_thickness}
B_\mathrm{tr}|_{B_N=0}=\frac{1}{2}\left(-B_\lambda+\sqrt{B_\lambda^2+4B_\mathrm{tr}^{0} B_\lambda\cos^2\theta}\right).
\end{equation}
 Here $B_\mathrm{tr}^0$ denotes the spin transition field for the case of zero-thickness, zero tilt angle, and no nuclear spin polarization. It equals $\frac{4}{9}(\frac{1}{g_e\xi})^2$ and $(\frac{1}{g_e\xi})^2$ for $\nu=2/3$ and $1/2$, respectively. Eq.\,(\ref{eq:Btr_finite_thickness}) is quite different from the $\cos^2\theta$ dependence expected for the zero-thickness 2DES (see Eq.\,\ref{eq:Btr_two3rds}). Hence, the measurements in  tilted fields can be used to evaluate how significant the finite thickness effect is.

\subsection{\label{subsec:NucSpins}Influence of nuclear spin polarization}
The Zeeman term in the Hamiltonian of the hyperfine interaction between an electron and the nuclei is
\begin{equation}
\mathcal{H}_\mathrm{Z}=A_\mathrm{HF}\hat{\mathbf{I}}_Z\cdot\hat{\mathbf{S}}_Z.
\end{equation}
The influence of the polarized nuclear spins on the electron Zeeman splitting can be described in terms of an effective field $B_N$
\begin{equation}
\mathcal{H}_\mathrm{Z}=g_e\mu_B B_N.
\end{equation}
For fully polarized nuclear spins in bulk GaAs $B_N$ equals $-5.3$\,T.~\cite{Paget77} The minus sign reflects that $\mathbf{B}_N$ is opposite to the external magnetic field $\mathbf{B}_\mathrm{tot}$. The nuclear spin polarizations of all three isotopes in GaAs (i.e.\ $^{69}$Ga, $^{71}$Ga and $^{75}$As) follow the Brillouin function. For small $B/T$ ($<0.5$\,Tesla/mK approximately), this function can be simplified to a Curie law form:
\begin{equation}
\label{eq:CurieLaw} \mathcal{P}_N=\frac{\langle I
\rangle}{I}=\frac{\gamma_{n}\hbar (I+1)B_{\rm tot}}{3k_{B}T}.
\end{equation}
For all nuclei in GaAs, $I$ equals $3/2$ and $\gamma_n$ is the nuclear gyromagnetic ratio.
Summing up the contributions from all three types of nuclei yields the following expression
\begin{equation}
\label{eq:BNthermal}
B_N=\sum\limits_{i=1}^{3}b_{N,i}\mathcal{P}_{N,i}\simeq\frac{0.87 \mathrm{mK}}{g_e}\frac{B_\mathrm{tot}}{T},
\end{equation}
where $b_{N,i}$ is the maximum effective field of the nuclei of type $i$. Note that $B_N$ is inversely proportional to the electron $g$-factor $g_e$. For a 2D electron system confined in a narrow GaAs quantum well, $|g_e|$ can be considerably smaller than the bulk value $|-0.44|$~\cite{Malinowski00}. It has been demonstrated in electron spin resonance experiments that strong  perpendicular magnetic fields can further decrease the magnitude of $g_e$~\cite{Dobers88b}.

Polarized nuclear spins only modify the electron Zeeman energy, i.e.\ $E_Z=g_e\mu_B(B_\mathrm{tot}+B_N)$. They do not affect the orbital motion of the electrons. Since the Coulomb energy remains unaltered, the spin transition fields for the $\nu=2/3$ and the $\nu = 1/2$ state change. Let's first consider the case of the $\nu=2/3$ spin transition when the sample is mounted perpendicular to the external applied magnetic field ($\theta=0$). The transition field is obtained from  equation
\begin{equation}
 \frac{\hbar e}{3 \xi m_e}\sqrt{\frac{B_\lambda B}{B+B_\lambda}} = g_e\mu_B(B+B_N).
\end{equation}
When $B_N$ is much smaller than $B$, one finds
\begin{equation}
\label{eq:Btr_smallBN}
B_\mathrm{tr}|_{\theta=0} \simeq B_\mathrm{tr}|_{\theta=0, B_N=0}-\left(1+\sqrt{\frac{B_\lambda}{B_\lambda+4B_\mathrm{tr}^0}}\right)B_N,
\end{equation}
where $B_\mathrm{tr}|_{\theta=0, B_N=0}=\frac{1}{2}\left(-B_\lambda+\sqrt{B_\lambda^2+4B_\mathrm{tr}^0}\right)$ is the transition field in the absence of  nuclear spin polarization. In the limit of zero-thickness, $B_\lambda\rightarrow\infty$, Eq.\,(\ref{eq:Btr_smallBN}) reduces to
\begin{equation}
B_\mathrm{tr}|_{\lambda=0, \theta=0} \simeq B_\mathrm{tr}^0-2B_N.
\end{equation}

 When the effect of $B_N$, the influence of finite thickness as well as a sample tilt are all included, the calculation of the transition
 field is more tedious. Nevertheless, a simplified treatment is possible when $B_N$ follows the Curie law. The transition field is obtained by solving
\begin{equation}
\frac{\hbar e}{3 \xi m_e}\sqrt{\frac{B_\lambda B}{B+B_\lambda}}=g_e\mu_B(1+\delta)B_\mathrm{tot},
\end{equation}
with $\delta=0.87\mathrm{mK}/(g_eT)$. This gives rise to a solution similar to Eq.\,(\ref{eq:Btr_finite_thickness}):
\begin{equation}
\label{eq:Btr_two3rds_full}
B_\mathrm{tr}=\frac{1}{2}\left(-B_\lambda+\sqrt{B_\lambda^2+4\frac{B_\mathrm{tr}^0}{(1+\delta)^2}B_\lambda\cos^2\theta}\right).
\end{equation}
For many quantum wells of interest, $g_e$ is smaller than zero and the nuclear spins in thermal equilibrium  increase the spin transition field. For a typical $g$-factor ($g_e\sim-$0.4), the correction due to nuclear spins is, however, rather small even at very low temperatures. For instance at $T=20$\,mK, $1+\delta\approx0.9$. Also noteworthy is that in the limit of zero-thickness, Eq.\,(\ref{eq:Btr_two3rds_full}) reduces to $B_\mathrm{tr}|_{\lambda=0}=B_\mathrm{tr}^0\cos^2\theta/(1+\delta)^2$.

\section{\label{sec:Methods}Experimental methods}

\subsection{\label{subsec:Measurement}Sample preparation and measurement setup}

The samples studied in this work are either single- or double-sided doped $\mathrm{Al}_x\mathrm{Ga}_{1-x}\mathrm{As}$/GaAs quantum wells with thicknesses between 16 nm and 18 nm.
Although qualitatively similar results were observed on all samples, the data presented here were recorded on a 16\,nm thick, single-sided doped quantum well  with an in-situ grown  $n^+$-GaAs backgate. The sample was patterned into 400\,${\rm \mu m}$ wide Hall bars with voltage probes along the top and bottom perimeter that are 400\,${\rm \mu m}$ apart. The electron density and mobility at zero backgate voltage are $1.77\times10^{11}$\,cm$^{-2}$ and $0.8\times10^6$\,cm$^{2}/$V$\cdot$s, respectively.

Electron transport measurements were carried out in dilution refrigerators with base temperatures less than $20$\,mK. For the tilted field measurements, the samples were mounted on a stage with
a low friction rotation mechanism driven by a high precision dc motor. The tilt angle $\theta$ was calibrated with low field
Hall measurements. Standard lock-in techniques were used for the transport measurements. In order to study the current induced effects, two currents are applied to the samples. The first is a small ac current which is typically 1\,nA, and the other is a dc current or an ac current with different frequency from the first one. The lock-in amplifiers are locked to the first ac current, and hence they measure the differential longitudinal resistance $dV_\mathrm{xx}/dI$ and the differential Hall resistance $dV_\mathrm{xy}/dI$, which are denoted as $R_\mathrm{xx}$ and $R_\mathrm{xy}$, respectively, for simplicity. The differential resistances could be considerably different from the longitudinal resistance $V_\mathrm{xx}/I$ and the  Hall resistance $V_\mathrm{xy}/I$ under sufficiently large bias current, but the latter are irrelevant for most of the discussion in this paper.


\subsection{\label{subsec:Two3rds}Nuclear magnetometry based on the $\nu=2/3$ spin transition}

\begin{figure}
\includegraphics*[width=7 cm]{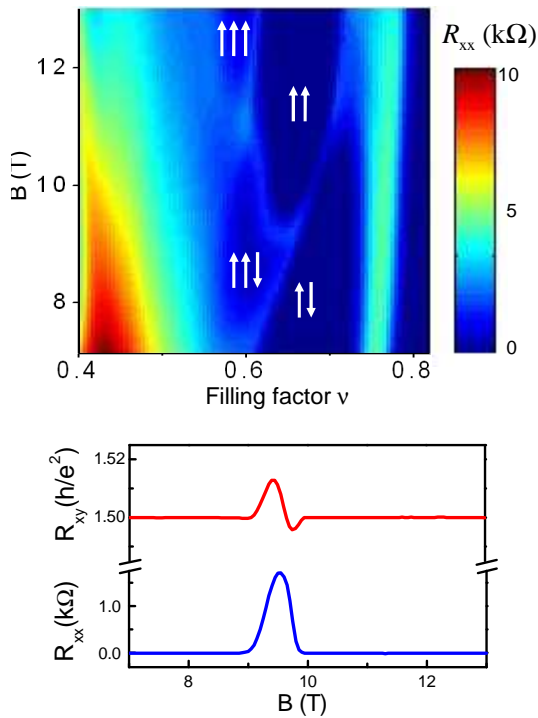}
\caption{\label{Fig2_Two3rd_Transition} Top panel: 2D plot of $R_\mathrm{xx}$ in the ($\nu,B$) plane with $I_\mathrm{ac}$=1\,nA at the base temperature. Regions labeled with $\uparrow \downarrow$ and $\uparrow \uparrow$ correspond to the incompressible $\nu = 2/3$ fractional quantum Hall state with spin polarization $\mathcal{P}=$0 and 1, respectively. The two ground states of the $\nu = 3/5$ quantum Hall fluid with different spin polarizations are also shown; Bottom panel: traces of $R_\mathrm{xx}$ and $R_\mathrm{xy}$ as a function of $B$ with $\nu$ fixed at 2/3. Dissipative transport takes place in the phase transition region.}
\end{figure}

As described in Sec.\,\ref{subsec:SpinTransition}, the crossing of the CF levels (0,$\downarrow$) and (1,$\uparrow$) is accompanied by a transition from an unpolarized ground state with $\mathcal{P}=0$ to a fully polarized ground state with $\mathcal{P}=1$. This spin transition as a result of the competition between the Coulomb energy and the Zeeman energy leaves a signature in a measurement of $R_\mathrm{xx}$, since longitudinal transport becomes dissipative near the transition~\cite{Engel92,Smet01,Kraus02,Hashimoto02,Hashimoto04}. Fig.\,\ref{Fig2_Two3rd_Transition} shows an example of a transport measurement. A color rendition of the longitudinal resistance in the filling factor versus $B$-field plane is depicted in the top panel. Arrows mark the spin orientation of the Landau levels which are occupied for states with fractional fillings 2/3 and 3/5. At filling factor 2/3 the resistance becomes non-zero
near 9.5 T. This is more clearly seen in a plot of the resistance along the line of constant filling factor 2/3 displayed in the bottom panel.
 Below this field the ground state is unpolarized. Also displayed in this graph is the Hall resistance. At the transition the Hall resistance  deviates from the quantized Hall resistance. At higher $B$-fields, the formation of the fully spin polarized ground state gives rise to reentrant behavior in $R_{\rm xx}$ and $R_{\rm xy}$. As described above, the transition field $B_\mathrm{tr}$ is determined by the relative strength of  $E_C$ and $E_Z$ and  can therefore be varied either by tilting the magnetic field~\cite{Kraus02}, or changing the nuclear spin polarization \cite{Smet02}. Conversely, a  measurement of the 2/3 transition field can serve as a method for nuclear magnetometry, since the displacement of the 2/3 transition provides information on the degree of nuclear spin polarization \cite{Smet04}.
It follows from Eq.\,(\ref{eq:Btr_smallBN}) that for small nuclear fields $B_N$, the change in $B_N$ is simply proportional to the shift of the transition along the magnetic field axis, or more specifically,
\begin{equation}
\label{eq:DeltaBN}
\Delta B_N \simeq -\left(1+\sqrt{\frac{B_\lambda}{B_\lambda+4B_\mathrm{tr}^0}}\right)^{-1}\Delta B_\mathrm{tr}.
\end{equation}

A very useful scheme to study the interaction physics between nuclear spins and electron spins at filling factors other than 2/3 was introduced in Ref~\cite{Smet04}. The measurement sequence is illustrated in Fig.\,3(a). It allows measuring the nuclear spin polarization at filling factor $\nu_\mathrm{rest}$, the filling factor of interest. Throughout this work, $\nu_\mathrm{rest}$ equals $1/2$.  The system is allowed to relax and reach a steady state during an extended period of time $t_\mathrm{rest}$ at filling $\nu_\mathrm{rest}$. In order to determine the degree of nuclear spin polarization at this filling factor, the magnetic field is swept in small steps in a range large enough to cover the 2/3 phase transition peak. Each time after changing the magnetic field slightly, the system is again allowed to relax during a time $t_\mathrm{rest}$. The gate voltage tracks the externally applied magnetic field to ensure that the electron system remains at $\nu_\mathrm{rest}$ even during the short $B$-field sweep. So the filling factor is at all times $\nu_\mathrm{rest}$ except during a short excursion period to $\nu_\mathrm{meas}$ (typically a value close to 2/3) where we perform the nuclear magnetometry. A small ac current ($I_\mathrm{ac}$ =1\,nA) is turned on for recording $R_\mathrm{xx}$ during this excursion time $t_\mathrm{meas}$.  In this work, $t_\mathrm{rest}$ was chosen to be 120 or 180\,s and $t_\mathrm{meas}$=1.5\,s in order to minimize the effect of nuclear spin relaxation at $\nu_\mathrm{meas}$ itself. It was verified that longer $t_\mathrm{rest}$ did not bring noticeable changes in the results on the nuclear spin phenomena at $\nu=1/2$. From the magnetic field at which the spin transition peak appears the effective field $B_{\rm N}$ is extracted. An important prerequisite to be able to calculate $B_\mathrm{N}$ is the knowledge of the transition field in the absence of nuclear spin polarization. How this reference value is obtained will be discussed in more detail in Sec.\,\ref{subsec:acNucDepol}.

\begin{figure}
\includegraphics*[width=8 cm]{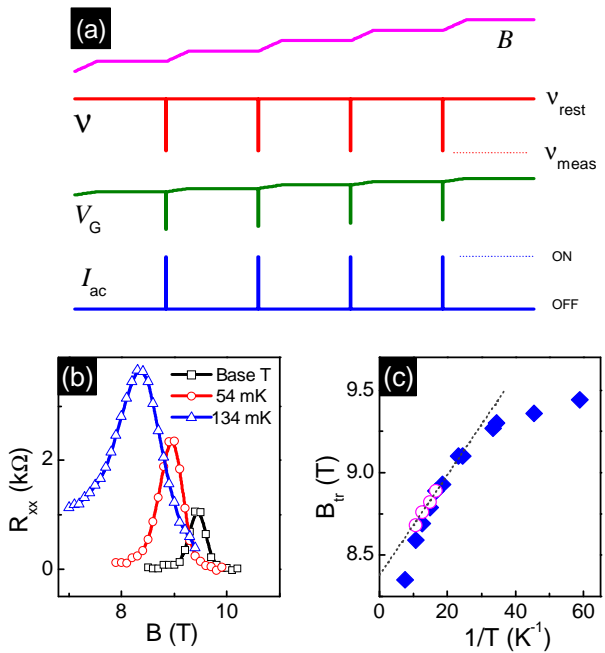}
\caption{\label{Fig3_Magnetometry_Two3rds}(a) Measurement sequence for nuclear magnetometry based on the spin transition at $\nu$=2/3. (b) 2/3 transition peaks for $\nu_\mathrm{rest}$=1/2 measured at $T$=17, 54, and 134\,mK. (c) Temperature dependence of the 2/3 spin transition field $B_\mathrm{tr}$ for $\nu_\mathrm{rest}$=1/2. Plotted in solid diamonds are $B_\mathrm{tr}$ values extracted from single peak fits. Open circles represent the corrected $B_\mathrm{tr}$ values in which the offsets from the effects irrelevant to nuclear spins are removed. The dotted line marks the expected Curie law dependence of the transition field. See Sec.\,\ref{subsec:acNucDepol} for more details.}
\end{figure}

Fig.\,3(b) displays $R_{\rm xx}$ traces recorded according to the sequence in panel a for three different temperatures. The spin transition at filling factor 2/3, signaled by the peak in $R_{\rm xx}$, moves to lower magnetic fields as $T$ increases. As discussed in the previous section, the partially polarized Fermi sea allows for efficient coupling between the nuclear and electron spin systems. At thermal equilibrium, the nuclear spin temperature is the same as the electron temperature. Cooling the electrons at $\nu=1/2$ increases the degree of nuclear spin polarization. The nuclear spin polarization acts back on the electron spins as a result of the reduced Zeeman energy, i.e.\ $E_Z=B_\mathrm{tot}+B_N$, with $B_N$ given in Eq.\,\ref{eq:BNthermal}. For the temperatures encountered in this experiment, $|B_N|<1$\,T, which is about one order of magnitude smaller than $B_\mathrm{tr}$, so the assumption under which Eq.\,(\ref{eq:DeltaBN}) has been derived is satisfied. One would expect that the shift on $B_\mathrm{tr}$ follows the Curie law and depends linearly on $1/T$. The data plotted in Fig.\,\ref{Fig3_Magnetometry_Two3rds}(c), however, clearly does not follow a $1/T$ behavior. We attribute this to two factors. The resistance maximum associated with the spin transition is broad and possesses an asymmetric background as a result of thermal activation at high temperatures. This precludes us to extract precise values for the transition fields. A second problem is the difficulty in determining the electron temperature. The actual electron temperature $T_e$ may deviate from the bath temperature $T$, even though the thermometer for measuring the bath temperature is mounted very close to the sample. The difference between $T_e$ and $T$ becomes non-negligible at temperatures lower than 45\,mK. This issue will be discussed in the following sections.

\begin{figure}
\includegraphics*[width=7.5 cm]{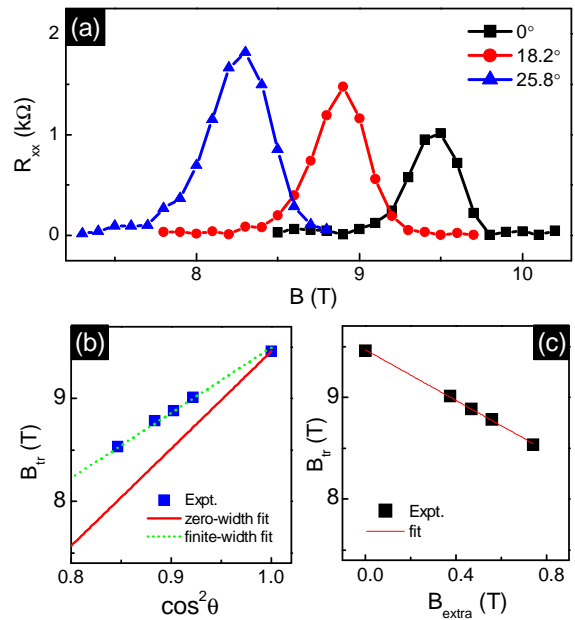}
\caption{\label{Fig4_Finite_thickness}(a) Measurement of the spin transition position at $\nu$=2/3 for $\nu_\mathrm{rest}=$1/2 in titled magnetic fields; (b) 2/3 transition fields plotted as a function of $\cos^2\theta$ (squares) and its fit to Eq.\,(\ref{eq:Btr_two3rds_full}) with finite thickness effect included (dotted lines). Also plotted for comparison is the expected angular dependence of $B_\mathrm{tr}$ for a zero-thickness 2DES (solid line); (c) $B_\mathrm{tr}$ as the function of $B_\mathrm{extra}$ (squares) and its linear fit (line), which can serve as a calibration curve for the nuclear magnetometry based on the 2/3 transition. The linear fit gives $\Delta B_\mathrm{tr} \approx -1.24\Delta B_N $ for the 16\,nm quantum well. This is  considerably different from $\Delta B_\mathrm{tr} \simeq -2\Delta B_N $ expected for the 2DES with zero-thickness.}
\end{figure}

The results above suggest that the dependence of the maximum in the spin transition peak on the bath temperature does not provide a reliable framework to extract the nuclear spin polarization based on the 2/3 spin transition. This difficulty can, however, be overcome by measuring the transition field $B_\mathrm{tr}$ in titled magnetic fields. In this work we limited the measurements to small tilt angles so that the orbital effect of the in-plane field can be ignored. As displayed in Fig.\,\ref{Fig4_Finite_thickness}(a), the peak of the 2/3 spin transition moves to lower perpendicular field as the tilt angle increases. The height of the peak increases considerably with tilt, but the width of the peak varies very little. This is in contrast with the temperature dependent behavior which is dominated by strong broadening at high temperatures. This feature is particularly helpful for a precise evaluation of $B_\mathrm{tr}$. The transition field is plotted in Fig.\,\ref{Fig4_Finite_thickness}(b) as a function of $\cos^2\theta$. The experimental values of $B_\mathrm{tr}$ deviate significantly from what one would expect for a zero-thickness 2D electron system. Fitting the data to Eq.\,(\ref{eq:Btr_two3rds_full}) yields $B_\mathrm{tr}^0/(1+\delta)^2\approx 26$\,T and $B_\lambda\approx$5.4\,T. The latter corresponds to an effective width of $\lambda\approx$11\,nm. It is slightly larger than the half width of the 16\,nm thick quantum well. Taking $(1+\delta)^{-2}\approx0.8$ at $T=20$\,mK (see Sec.\,\ref{subsec:NucSpins}), one obtains $B_\mathrm{tr}^0\approx$21\,T. It follows from Eq.\,(\ref{eq:Btr_smallBN}) that $\Delta B_\mathrm{tr} \approx -1.2\Delta B_N$, significantly different from the $\Delta B_\mathrm{tr}=-2B_N$  expected in the limit of zero-thickness.

The main effect of the small angle tilted field is that the extra Zeeman energy brought by the in-plane field lowers the transition field. This is very similar to the role of thermally depolarized nuclear spins. Therefore, the extra Zeeman field, defined as $B_\mathrm{extra}=B_\mathrm{tr}(1/\cos\theta-1)$, could provide a convenient route to determine $\Delta B_N$ without the need for estimating $B_\mathrm{tr}^0$. An example  is plotted in Fig.\,\ref{Fig4_Finite_thickness}(c). The linear fit of the $B_\mathrm{tr}$-$B_\mathrm{extra}$ data gives $\Delta B_\mathrm{tr}\approx -1.24 B_\mathrm{extra}$, or equivalently $\Delta B_\mathrm{tr} \approx -1.24\Delta B_N $ in case that the nuclear spins are involved.

\subsection{\label{subsec:OneHalf}Nuclear magnetometry based on  the $\nu=1/2$ spin transition}

The nuclear magnetometry based on the $\nu=$2/3 spin transition requires a rather sophisticated measurement sequence. Its application is limited to a narrow magnetic field range in the vicinity of  the $\nu = 2/3$ spin transition only. For a 2D electron system residing in a wider quantum well, which is desirable in order to benefit from higher electron mobilities, the 2/3 transition peak does not even show up in the transport measurement at sufficiently low $T$. This makes the $\nu=$2/3 detection scheme no longer useful. Moreover, the magnetic field sweeps during the measurement sequence raise concerns over whether the nuclear spin system is truly in thermal equilibrium. This becomes even more problematic for filling factors at which the coupling between nuclear and electron spins is weak.

\begin{figure}
\includegraphics*[width=7.5 cm]{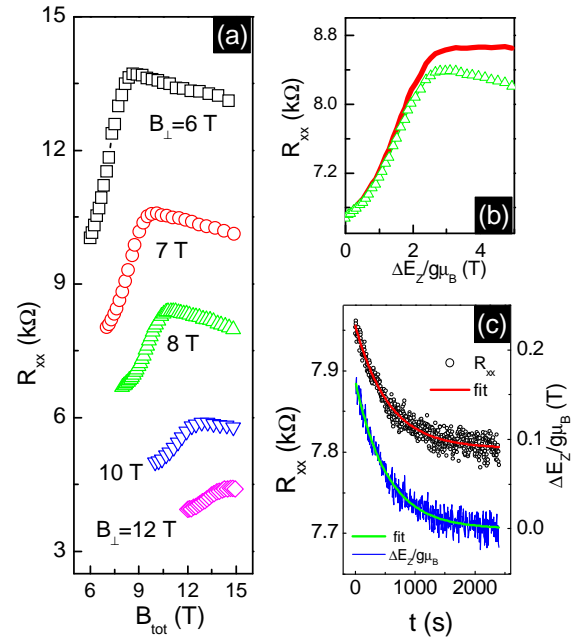}
\caption{\label{Fig5_OneHalf_Magnetometry}(a) Longitudinal resistance $R_\mathrm{xx}$ at $\nu=1/2$ as a function of $B_\mathrm{tot}$ with the perpendicular field $B$ fixed at 6\,T, 7\,T, 8\,T, 10\,T, and 12\,T. (b) $R_\mathrm{xx}$ plotted as a function of the extra Zeeman field. Triangles are experimental points obtained by directly converting $B_\mathrm{tot}$ to $B_\mathrm{tot}-B$. The line is the result of subtracting the contribution from the orbital effect of the in-plane magnetic field. (c) An example of the time evolution of $R_\mathrm{xx}$ (open dots) as a consequence of nuclear spin relaxation at $\nu=1/2$  and the corresponding change in the Zeeman field $\Delta E_Z/(g_e\mu_B)$ (line) as a function of time. The conversion from $R_\mathrm{xx}$ to $\Delta E_Z/(g_e\mu_B)$ is based on the calibration data  in (b). }
\end{figure}

Here we describe a new type of nuclear magnetometry. It is based on the Fermi sea at $\nu=1/2$. The degree of  spin polarization of this Fermi sea is also determined by $E_Z/E_C$ and can therefore readily be tuned as well by either changing the nuclear spin polarization~\cite{Tracy07}, or by tilting the sample while keeping the perpendicular field $B$ fixed~\cite{Li09}. For the 16\,nm thick sample used in this work, the electron system remains partially polarized up to at least 16\,T if the magnetic field is not tilted. In tilted field measurements with a fixed perpendicular field $B$, the longitudinal resistance $R_\mathrm{xx}$ increases with $B_\mathrm{tot}$ (or $E_Z$) until it reaches a maximum  near full spin polarization. As demonstrated in our previous work~\cite{Li09}, $R_\mathrm{xx}$ no longer responds to the change in $E_Z$ in case full spin polarization has been reached. The $R_\mathrm{xx}$-$B_\mathrm{tot}$ curves in Fig.\,\ref{Fig5_OneHalf_Magnetometry}(a) can be interpreted as $R_\mathrm{xx}$ versus $E_\mathrm{Z}$ curves and hence changes in the resistance can be converted into changes of the nuclear field, i.e. the degree of nuclear spin polarization. The orbital effect of the in-plane field, which is responsible for the small negative slope at large $B_\mathrm{tot}$, should be subtracted for large tilt angles. Fig.\,\ref{Fig5_OneHalf_Magnetometry}(b) shows an example for $B=8$\,T.

Nuclear magnetometry using the properties of the CF Fermi sea at $\nu=1/2$ is performed with the following sequence of operations: First, the sample relaxes at $\nu=1/2$, usually with a small current (typically 1\,nA) applied for monitoring $R_\mathrm{xx}$. The relaxation time, $t_\mathrm{relax}$, is usually chosen about one order of magnitude longer than the nuclear spin relaxation time $T_1$. In this experiment, $t_\mathrm{relax}$=900\,s unless otherwise specified. The nuclear spin system is expected to be close to equilibrium with the electron spin system at the end of this time period. The sample is then brought to the state of interest at filling factor $\nu_{pol}$ for a certain time. This may be the same or a different filling factor and RF radiation may be turned on in resonance with nuclear spins or the system may also be excited by a large current. Subsequently, the filling factor is set back to 1/2 (if it has been changed in the previous step) and  all of the external excitation sources (if any were turned on) are shut off. Only the small measurement current ($I_\mathrm{ac}=$1\,nA) remains turned on in order to record the relaxation of $R_\mathrm{xx}$ at filling 1/2. After $R_\mathrm{xx}$ has saturated, the system is ready for a new set of measurements.

The time evolution of $R_\mathrm{xx}$ at filling 1/2 recorded after the excursion to filling factor $\nu_\mathrm{pol}$ can be fitted to the following exponential decay function:
\begin{equation}
\label{eq:ExpDecayRxx}
R_\mathrm{xx}(t)=R_0+\Delta R\exp(-t/\tau).
\end{equation}
Comparing $\Delta R$ with the $R_\mathrm{xx}$ curve recorded at a fixed perpendicular magnetic field as a function of tilt angle (for instance the one shown in Fig.\,\ref{Fig5_OneHalf_Magnetometry}(b)) enables to extract the time dependent change in the Zeeman energy, and hence the change in $B_N$. It should be noted that the time constant $\tau$ is equal to $T_1$ only when $R_\mathrm{xx}$ depends linearly on $E_Z$. A strong non-linearity would cause a large discrepancy between $\tau$ and $T_1$. A fitting procedure applicable even if $R_\mathrm{xx}$ depends in a non-linear fashion on $E_\mathrm{Z}$  can however be easily devised by converting the $R_\mathrm{xx}$ values to values of $\Delta E_Z/(g_e\mu_B)$ and then fitting the data to
\begin{equation}
\label{eq:ExpDecayDeltaEz}
\Delta E_Z(t)=\Delta E_Z^0\exp(-t/T_1).
\end{equation}
 Fortunately for much of the region where the $\nu=1/2$ state is partially polarized, a linear approximation is justified if  $\Delta E_Z$ remains small and consequently $T_1$ is usually close to $\tau$. For example, the relaxation shown in Fig.\,\ref{Fig5_OneHalf_Magnetometry}(c) yields a $\tau=250$\,s whereas the procedure using Eq.\,(\ref{eq:ExpDecayDeltaEz}) gives $T_1=255$\,s.

\section{\label{sec:Result}Results and Discussion}

\subsection{\label{subsec:NucDepol}Current induced nuclear spin depolarization detected with the $\nu=2/3$ spin transition}

\begin{figure}
\includegraphics*[width=7.5 cm]{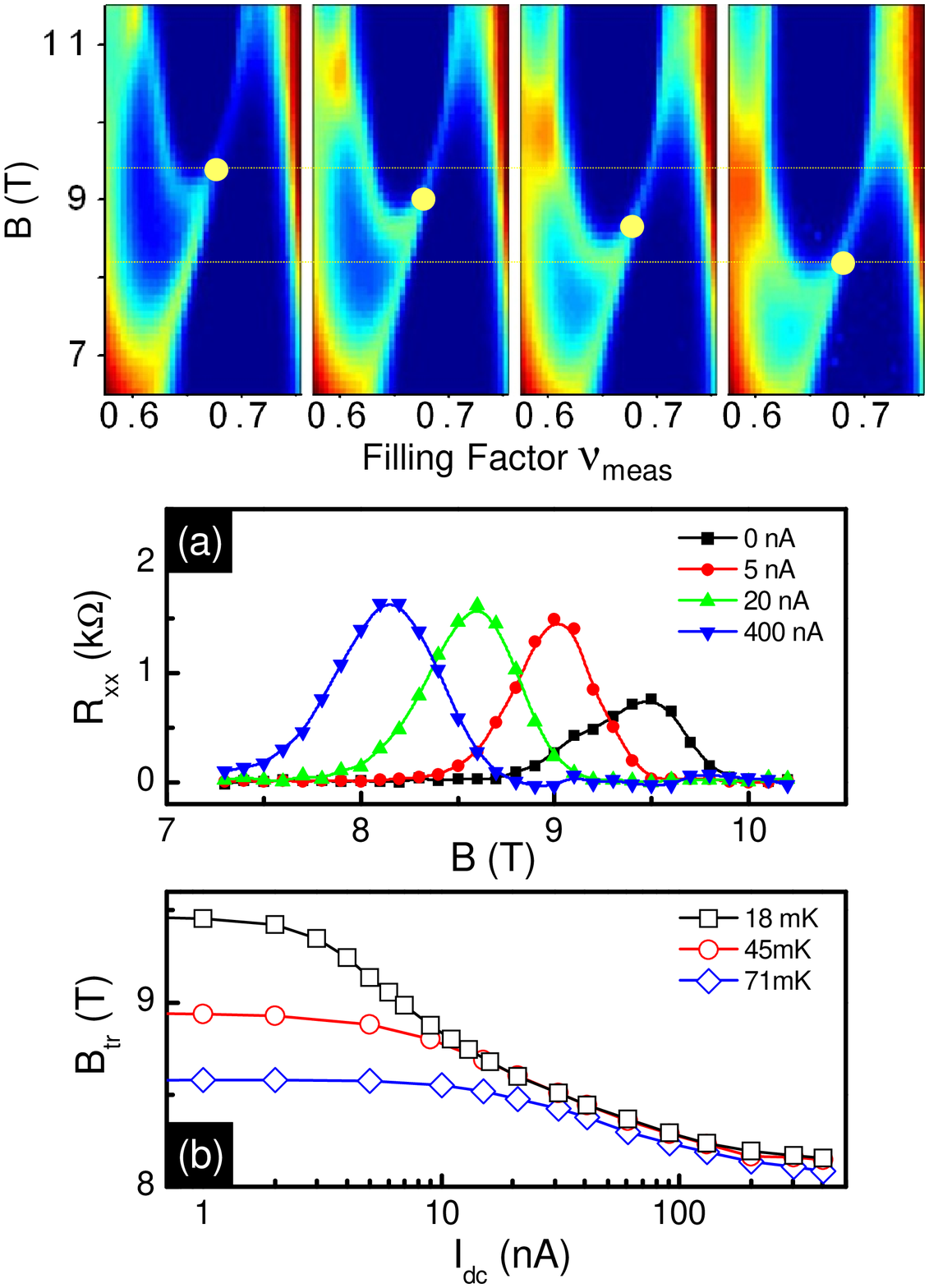}
\caption{\label{Fig6_NucSpinDepol_dc} Top panels: The phase transition diagrams of $R_\mathrm{xx}$ (defined as $\frac{dV_\mathrm{xx}}{dI}$) plotted in the ($\nu_\mathbf{meas}$, $B$) plane with $\nu_\mathrm{rest}$=1/2 and an applied $I_\mathrm{dc}$ of 0, 5, 20, 400\,nA from left to right. The detection filling factor $\nu_\mathrm{meas}$ is varied from 0.575-0.755 for each of the diagrams. Bottom panels:
(a) The $\nu=2/3$ spin transition peaks for an applied $I_\mathrm{dc}$=0, 5, 20, 400\,nA during the time the system is left at $\nu_{\rm rest}$=1/2; (b) The dc current dependence of the phase transition field $B_\mathrm{tr}$ at $T$=18, 45, and 71\,mK. See text for details.}
\end{figure}

Fig.\,\ref{Fig6_NucSpinDepol_dc} illustrates measurements performed by using the sequence described in Sec.\,\ref{subsec:Two3rds} involving the spin transition at filling 2/3. These experiments were carried out in order to investigate how the dc current influences the nuclear spin polarization at filling 1/2. For each filling factor $\nu_\mathrm{meas}$, the magnetic field was swept from 6.5\,T to 11.5\,T in steps of 0.1\,T. The field sweep rate was 0.1\,T/minute. The field was then set to remain constant for 2\,minutes before being ramped to the next value. The filling factor was fixed at 1/2 all the time except for a short excursion period of 1.5\,s during which the filling factor was changed to $\nu_\mathrm{meas}$ and a small ac current ($I_\mathrm{ac}$=1\,nA) was switched on to record $R_\mathrm{xx}$. The dc current $I_\mathrm{dc}$ was turned on only when $\nu_\mathrm{rest} = 1/2$. Repeating the above measurement for $\nu_\mathrm{meas}$=0.575 to 0.755 resulted in the color renditions of $R_\mathrm{xx}$ in the ($\nu_\mathrm{meas}$,$B$)-plane.

The four color plots in Fig.\,\ref{Fig6_NucSpinDepol_dc} (top panels) reveal the spin phase transition for the $\nu = 2/3$ state at base temperature when a dc current excitation is applied during the time the system is kept at $\nu_\mathrm{rest} = 1/2$ with $I_\mathrm{dc}=$0, 5, 20 and 400\,nA (from left to right). The phase transition moves to lower field values as $I_\mathrm{dc}$ is increased.  The transition field shifts by more than  1\,T when $I_\mathrm{dc}$ is varied from 0 to 400\,nA.  Fig.\,\ref{Fig6_NucSpinDepol_dc}(b) shows the dependence of the transition field $B_\mathrm{tr}$ on $I_\mathrm{dc}$ for three different bath temperatures $T=$18, 45  and 71\,mK. As $T$ becomes higher, $B_\mathrm{tr}$ remains independent of the dc current up to high current values. At higher dc currents however, the curves nearly coincide.

\begin{figure}
\includegraphics*[width=7.5 cm]{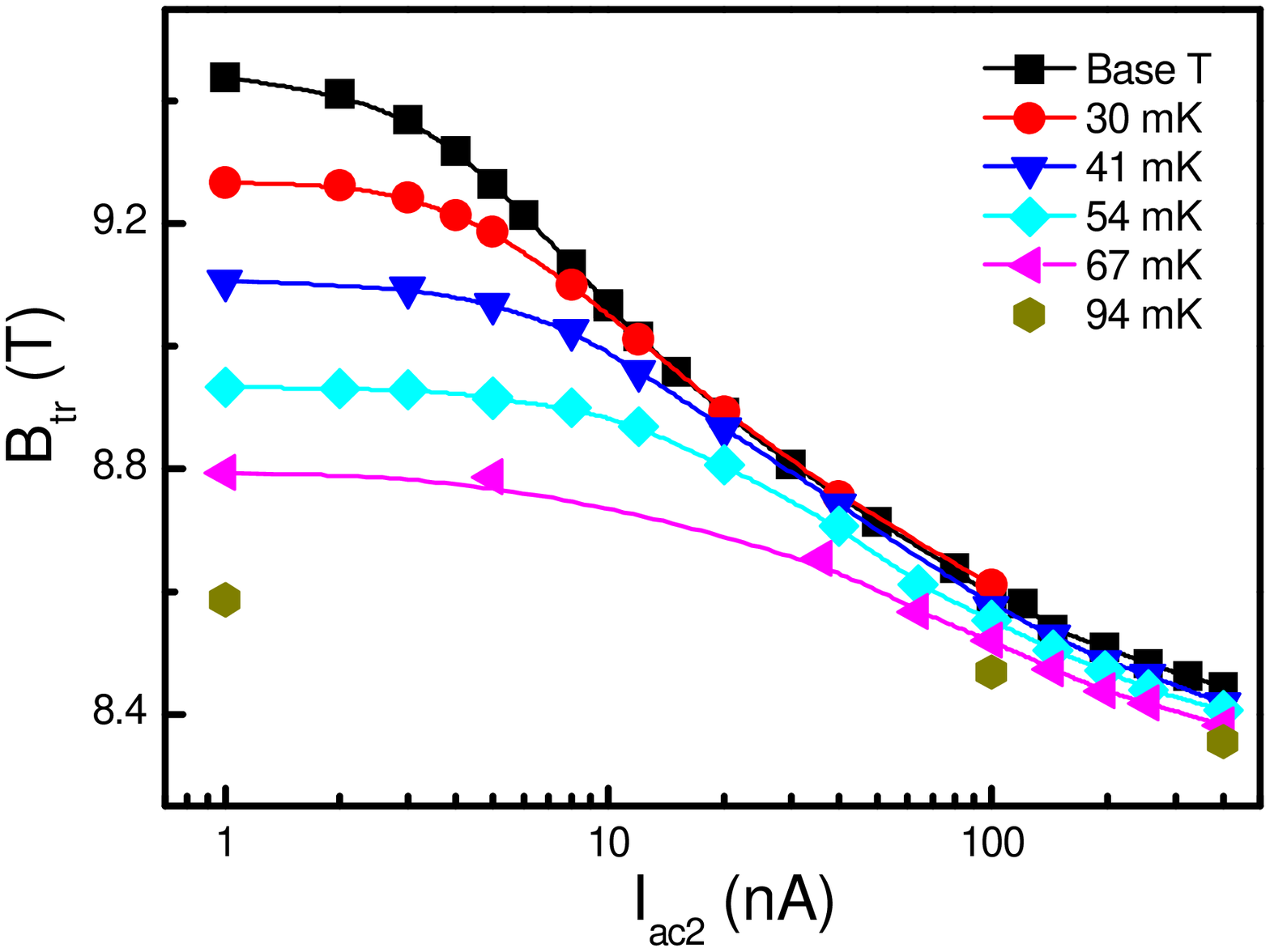}
\caption{\label{Fig7_NucSpinDepol_ac} The ac current dependence of the phase transition field $B_\mathrm{tr}$ for $\nu_\mathrm{rest}$=1/2 at various bath temperatures. The values of $B_\mathrm{tr}$ are extracted from the spin transition peaks of $R_\mathrm{xx}$ with $I_\mathrm{ac}=1$\,nA. The lock-in amplifier is locked to $I_\mathrm{ac}$. The frequency of the second ac current $I_\mathrm{ac2}$ is chosen to be different from $I_\mathrm{ac}$ so that it does not interfere with the measurement of the differential resistance.}
\end{figure}

Similar effects were also observed for ac current excitation. Fig.\,\ref{Fig7_NucSpinDepol_ac} shows the results of a set of measurements carried out with a similar measurement sequence, except that $I_\mathrm{dc}$ is replaced with an ac current, denoted as $I_\mathrm{ac2}$ throughout this paper, with a frequency different from the one for measuring $R_\mathrm{xx}$, i.e.\ $I_\mathrm{ac}$. As previously for the dc-current, $I_\mathrm{ac2}$ was turned on during the time period  when the filling is set to $\nu_\mathrm{rest} = 1/2$.

\subsection{\label{subsec:CIERxx}Current induced effect on electron transport properties}

In Sec.\,\ref{subsec:Two3rds}, we have learned that raising the temperature lowers $B_\mathrm{tr}$ as a result of nuclear spin depolarization. The current induced decrease in $B_\mathrm{tr}$ is also attributed to changes in the nuclear spin polarization.  A decrease in $B_\mathrm{tr}$ corresponds to a positive $\Delta B_N$ or a decrease of $|B_N|$. It is natural to suspect that the electron system gets heated as a consequence of weak electron-phonon coupling at ultra-low temperatures and that the entropy is transferred to the nuclei in case  a strong interaction between the electron spins and nuclear spins exists.  The current induced electron heating has been detected previously in many systems~\cite{Dorlan79,Roukes85,Chow96}, but to the best of our knowledge, the electron heating effects at $\nu=1/2$ have not been studied systematically.
At $\nu=1/2$, the nuclear spin system near the 2DES has only one, but very effective way interacting with the environment, namely the spin flip-flop process via the Fermi contact hyperfine interaction. The elevated temperature of the electron system due to the applied current can therefore be transferred to the nuclear spin system.


\begin{figure}
\includegraphics*[width=7.5 cm]{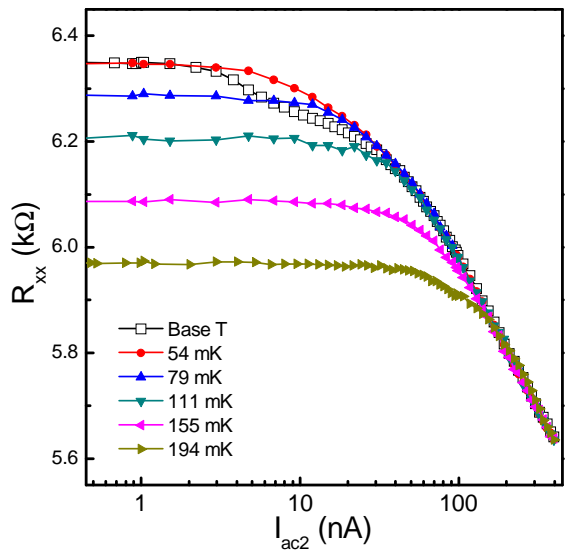}
\caption{\label{Fig8_CIE_transport} The current induced effects on transport properties at $\nu$=1/2 measured at the various bath temperatures (less than 20 to 194\,mK). The differential resistance $R_\mathrm{xx}$ is plotted as a function of the second ac current $I_\mathrm{ac2}$. A perpendicular magnetic field of $B$=8\,T was applied in all of the measurements.}
\end{figure}

The longitudinal resistance $R_\mathrm{xx}$ is subject to temperature dependent quantum corrections at low temperatures. This temperature dependence of $R_\mathrm{xx}$ seems the most obvious route to estimate the actual electron temperature under the influence of an externally imposed current in order to verify the above explanation for the current induced decrease of $B_\mathrm{tr}$ or $|B_N|$. A similar approach to estimate the electron temperature has been used in other systems such as for example copper thin films~\cite{Roukes85}. The ac current dependence of $R_\mathrm{xx}$ plotted in Fig.\,8 was measured as follows. Two ac currents were applied. One is $I_\mathrm{ac}$=1\,nA with a frequency of 22.7\,Hz, which was locked to the detection electronics for recording $R_\mathrm{xx}$. The frequency of the other current, $I_\mathrm{ac2}$, was set to 83.9\,Hz. At $T>54$\,mK, the overall behavior of the current induced changes in the longitudinal resistance is similar to those observed previously in   copper thin films~\cite{Roukes85}. Yet, in our system there are  some complications that limit the usefulness for extracting the actual electron temperature at lower temperatures. $R_\mathrm{xx}$ exhibits an anomaly at $\nu=$1/2 seen in Fig.\,8. The resistance at the base temperature with $I_\mathrm{ac2}\sim$10\,nA applied is  lower than that at $T=$54\,mK. It is opposite to what one would expect from the electron heating picture.
This is not completely surprising if one considers that the strongly interacting Fermi sea of electrons at filling 1/2 is not an ordinary Fermi liquid~\cite{HLR93}. No theory is available yet to describe the transport properties of the partially polarized Fermi sea at 1/2. Sizeable nuclear spin polarization at low $T$ further complicates the interpretation of this nonlinear temperature dependence of $R_\mathrm{xx}$. Because of these complications, we did not further pursue  the temperature dependence of $R_\mathrm{xx}$ to evaluate the amount of current induced electron heating in this work.

\subsection{\label{subsec:T1thermometry}Using $T_1$ for sensing the electron temperature}

\begin{figure}
\includegraphics*[width=7.5 cm]{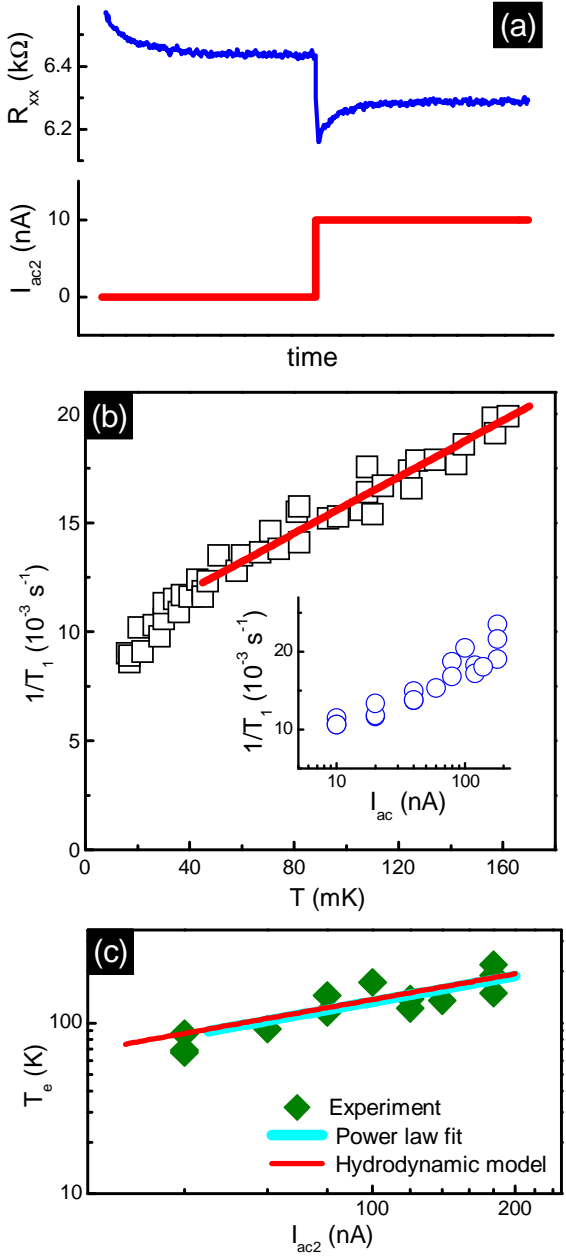}
\caption{\label{Fig9_T1_Thermometry}(a) An example of the measurement sequence for extracting the nuclear spin relaxation rate $1/T_1$. A small ac current, $I_\mathrm{ac}$ is always turned on to measure the (differential) resistance $R_\mathrm{xx}$, and the second ac current $I_{\rm ac2}$ (usually much larger than $I_\mathrm{ac}$) is turned on and off alternately to drive the nuclear spin polarization into different values. An averaging over multiple runs and small changes in the nuclear spin polarization were used in the measurements in order to obtain reliable $1/T_1$-data. (b) Temperature dependence of $1/T_1$. The squares are the experimental data and the solid line is a linear fit down to $\sim 45$\,mK. In the lower inset $1/T_1$ at the base temperature is plotted as a function of $I_\mathrm{ac2}$. (c) $I_\mathrm{ac2}$ dependence of the electron temperature $T_e$ (diamonds). Also plotted for comparison are the fits to the square root of the current, i.e.\ $T_e\propto\sqrt{I_\mathrm{ac2}}$ (thicker line), as well as the predicted values of the parameter-free hydrodynamic model (thinner, red line).}
\end{figure}

Here we describe a more attractive alternative to sense the actual electron temperature using the nuclear spin relaxation rate.
At filling factor 1/2, efficient coupling between nuclear spins and electron spins is assured in case of partial spin polarization of the electronic system, because of the continuous energy spectrum near the Fermi energy for both electron spin directions. For a disorder free and non-interacting composite fermion system at $\nu=$1/2, the nuclear spin relaxation is described by the Korringa law. It states that the inverse of the relaxation time, $1/T_1$, is proportional to $D_\uparrow(\varepsilon_F) D_\downarrow(\varepsilon_F) T$. Here,   $D_{\uparrow,\downarrow}(\varepsilon_F)$ are the density of states at the Fermi energy for up and down spins, respectively. If the spin polarization is close to  100\%, disorder will result in an inhomogeneous spatial distribution of the minority spins and deviations from the Korringa law have been observed~\cite{Li09}. Nevertheless, $T_1$ still depends monotonically on temperature, as discussed below and shown in Fig.\,\ref{Fig9_T1_Thermometry}(b).

A measurement of $T_1$ in the presence of an extra ac current $I_{\rm ac2}$ was carried out as follows. An ac current of $I_\mathrm{ac}=$1-10\,nA was applied to the sample to monitor the time dependence of $R_\mathrm{xx}$. The system was left to rest for 900\,s before the second ac current, $I_\mathrm{ac2}$, was turned on. The resistance as a function of time is plotted in Fig.~9(a). It changes immediately after turning on $I_{\rm ac2}$ due to heating of the electronic system, The resistance would then change as a function of time as a result of the current induced agitation of the nuclear spins. This time dependent data after turning on $I_{\rm ac2}$ was fitted to the exponential decay function [Eq.\,(\ref{eq:ExpDecayRxx})] to extract $T_1$. The result is displayed in the lower inset of Fig.\,\ref{Fig9_T1_Thermometry}(b). The nuclear spin relaxation rate $1/T_1$ increases as $I_{\rm ac2}$ becomes larger. This is presumably due to the electron heating. Being aware that the bath temperature may differ from the electron temperature at very low $T$, we only fit the $1/T_1$-$T$ data down to 45\,mK in order to obtain a reliable curve to extract the electron temperature $T_e$ from the $1/T_1$ values. The $1/T_1$-$I_\mathrm{ac2}$ data is then converted. $T_e$ is determined for every value of  $I_\mathrm{ac2}$ by using the $1/T_1$ versus $T$ data. The outcome of this conversion is depicted in Fig.\,\ref{Fig9_T1_Thermometry}(c). The electron temperature approximately follows the power law $T_e\approx13.3\sqrt{I_\mathrm{ac2}/\mathrm{nA}}$\,mK. This is very close to the behavior predicted by the hydrodynamical model, which was developed long ago to describe electron heating in the plateau-to-plateau transition region in quantum Hall systems~\cite{Chow96}. According to this model, $T_e\simeq 24.9(\sigma_\mathrm{xx}\rho_\mathrm{xx})^{1/4}[J/\mathrm{(A/m)}]^{1/2}$\,K, where $J$ is the current density. This would give $T_e\approx13.7\sqrt{I_\mathrm{ac2}/\mathrm{nA}}$\,mK. It is noteworthy that there is no free parameter in the hydrodynamic model~\cite{Chow96}. Hence, from the nuclear spin relaxation rate we can determine the actual electron temperature.

\subsection{\label{subsec:acNucDepol}Mechanism for current induced nuclear spin depolarization}

\begin{figure}
\includegraphics*[width=7.5 cm]{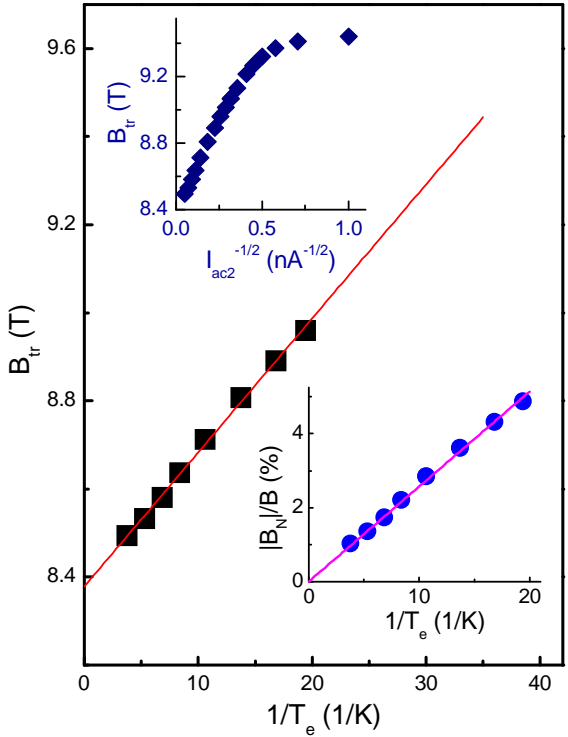}
\caption{\label{Fig10_Btr_Te_Dep} The $\nu=2/3$ transition field $B_\mathrm{tr}$ for $\nu_\mathrm{rest}$=1/2 plotted as a function of electron temperature $T_e$. The upper inset shows the raw data of the $I_\mathrm{ac2}$ dependence of $B_\mathrm{tr}$ measured at the base temperature. The electron temperature is converted from $I_\mathrm{ac2}$ using the thermometry based on the nuclear spin relaxation rates shown in Fig.\,\ref{Fig9_T1_Thermometry}.  In the lower inset the corresponding $|B_N|/B$, namely the ratio between the magnitude of the nuclear field and external magnetic field is plotted as function of $1/T_e$. It follows the Curie law $B_N= \frac{0.87\,\mathrm{mK}}{g_e}\frac{B}{T_e}$.}
\end{figure}

The extraction of the electron temperature from the nuclear spin relaxation measurements is helpful to gain insight into the influence of the current on the nuclear spin polarization. We illustrate this with the data recorded in Fig.\,7 for the position of the spin phase transition at filling factor 2/3 at base temperature as a function of the applied current $I_\mathrm{ac2}$. In the top inset of Fig.\,\ref{Fig10_Btr_Te_Dep}, this raw data has been replotted using $1/\sqrt{I_\mathrm{ac2}}$ as abscissa in view of the close connection between $1/\sqrt{I_\mathrm{ac2}}$ and $1/T_\mathrm{e}$. In the main graph, the current has been converted into the electron temperature using the nuclear spin relaxation data of Fig.~9.  Data points are only shown for those points for which the electron temperature is no longer determined by the bath temperature, but predominantly controlled by the applied $I_\mathrm{ac2}$. The 2/3 spin transition field $B_\mathrm{tr}$ follows a linear dependence on $1/T_e$, namely $B_\mathrm{tr}=(8.38+0.0305/T_e$)\,T. For the smallest $I_{\rm ac2}$, $B_\mathrm{tr}$ reaches $9.44\,T$ (top inset). From the linear fit, we conclude that the electron temperature  $T_e$ equals $28.7\,\mathrm{mK}$ in this case, which is considerably higher than the base temperature of the bath (15-18\,mK).

Based on the calibration of the finite thickness effect obtained in tilted field measurements, i.e.\ $\Delta B_\mathrm{tr}\simeq -1.24B_N$ (see Sec.\,\ref{subsec:Two3rds}), the $B_\mathrm{tr}$ data can be converted into $B_N$ and are displayed in the lower inset of Fig.\,\ref{Fig10_Btr_Te_Dep}. Since $B_{\rm N}$ can be written as $0.87 \mathrm{mK} B / (g_\mathrm{e} T_\mathrm{e})$ according to Eq.\,(\ref{eq:BNthermal}), it is possible to extract the electron $g$-factor $g_e\approx-0.34$  for this 16\,nm thick quantum well from this slope. This is consistent with a previous ESR experiment, in which the electron $g$-factor of a 15\,nm thick GaAs quantum well was determined to be $g_e=-(0.40-0.00575*B)$ for the lowest Landau level~\cite{Dobers88b}. At $B=$9.44\,T, the ESR experiment would give $g_e=-0.35$. Considering there is about 5\% uncertainty in the evaluation of $T_e$ from the nuclear relaxation time $T_1$, the agreement in the $g$-factor with ESR experiments is good. The linear fit of the $1/T_e$ dependence of the transition field also leads to $B_\mathrm{tr}=8.38$\,T in the limit $B_N\rightarrow 0$. An effective nuclear field $B_N=-(9.44-8.38)/1.24\approx-0.85$\,T at base temperature or an electron temperature $T_e = 28.7\ {\rm mK}$ can therefore be deduced from these measurements.

\begin{figure}
\includegraphics*[width=7.5 cm]{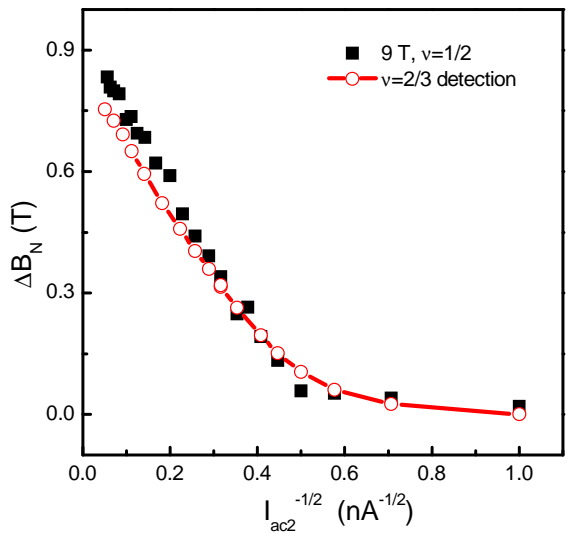}
\caption{\label{Fig11_Compare_OneHalf_Two3rds} Comparison of the two methods of nuclear magnetometry, which are based on the spin transitions at $\nu$=2/3 (open symbols) and $\nu$=1/2 (solid squares). The ac current ($I_\mathrm{ac2}$) induced change in nuclear field, $\Delta B_N$, at $\nu$=1/2 is plotted as a function of $I_\mathrm{ac2}^{-1/2}$, which is proportional to $1/T_e$. See text for details.}
\end{figure}

An independent confirmation of the validity of the nuclear magnetometry based on the 2/3 spin phase transition comes from the resistive detection method at $\nu=1/2$. The measurement sequence has been described in detail in Sec.\,\ref{subsec:OneHalf}, but is repeated here briefly for the sake of clarity. The system is allowed to equilibrate for $t_\mathrm{relax} = 900 s$ at half filling in the presence of a small current $I_\mathrm{ac} = 1nA$ used to monitor $R_\mathrm{xx}$, then $I_\mathrm{ac2}$ is turned on to depolarize the nuclear spins for 900 s. This depolarizing current is turned off and the time dependence of $R_\mathrm{xx}$ is recorded during a time period $t_\mathrm{relax}$. The procedure is then repeated for different values of $I_\mathrm{ac2}$.  The $R_\mathrm{xx}$ relaxation data during timeperiod $t_\mathrm{relax}$ can then be fitted to Eq.\,(\ref{eq:ExpDecayRxx}) in order to extract $\Delta B_N$ as described in Sec.\,\ref{subsec:OneHalf}. The data are summarized  in Fig.\,\ref{Fig11_Compare_OneHalf_Two3rds}. The change in the nuclear field, $\Delta B_N$, again has a linear dependence on $1/\sqrt{I_\mathrm{ac2}}$ (or $T_e$), similar to that observed in the $\nu$=2/3 detection experiment, in the regime where the electron temperature is controlled by the externally imposed current $I_\mathrm{ac2}$ and not by  the bath temperature $T$. Extrapolating the data to the high current limit (corresponding to $T_e\rightarrow \infty$), one obtains $\Delta B_N\approx0.9$\,T for $B=$9\,T or a degree of nuclear spin polarization corresponding to $B_N = -0.9\ {\rm T}$ at base temperature. This is close to $B_N\approx-0.85$\,T extracted from the nuclear magnetometry method based on the 2/3 transition.

\subsection{The Curie law dependence}

The Curie law dependence of $B_\mathrm{tr}$ on $T_e$ obtained from the current induced nuclear spin depolarization data is also consistent with the $B_\mathrm{tr}$ values measured with $I_\mathrm{ac2}$=0 in the experiment described in section III.B and summarized in Fig.\,\ref{Fig3_Magnetometry_Two3rds}, where the temperature of the bath is tuned in the intermediate temperature regime $T\sim$40-80\,mK. Deviations from the Curie law at higher temperatures are related to the shape of the 2/3 transition peak as a result of the thermally activated electron transport. It broadens the peak and makes it asymmetric which prevents a reliable extraction of $B_\mathrm{tr}$. This difficulty can be overcome, however, by using the information obtained from our studies on the influence of an additional ac-current $I_{\rm ac2}$. As shown in Fig.\,\ref{Fig8_CIE_transport}, for large enough $I_\mathrm{ac2}$,  the longitudinal resistivity at $\nu=1/2$ is independent of the bath temperature. In this case, the electron temperature  only depends on the current $I_{\rm ac2}$, even though it may not be the only parameter that controls the transport properties, as discussed in Sec.\,\ref{subsec:CIERxx}. The small temperature dependence of $B_\mathrm{tr}$ observed at large $I_\mathrm{ac2}$  in Fig.\,\ref{Fig7_NucSpinDepol_ac} can be attributed to artefacts related to the thermal broadening and asymmetric shape of the transition peak at high temperatures. Without these effects, all of the $B_\mathrm{tr}$-$I_\mathrm{ac2}$ data in Fig.\,\ref{Fig7_NucSpinDepol_ac} would merge into  a single curve at sufficiently high current densities. The offsets in $B_\mathrm{tr}$ observed in the large current limit can therefore be used to correct the $B_\mathrm{tr}$ values at high temperatures for $I_\mathrm{ac2}=0$. As shown in Fig.\,\ref{Fig3_Magnetometry_Two3rds}(c), the corrected $B_\mathrm{tr}$ values (open circles) agree very well with the Curie law dependence deduced from the current dependence of $B_\mathrm{tr}$ measured at the base temperature.

Now that all aspects of the experimental data are consistent with each other,  we are in a position to extract the effective mass of the composite fermions with the help of Eq.\,(4-13). Taking $T_e=28.7$\,mK and $g_e=-0.34$, we obtain $1+\delta\approx0.87/|g_eT_e|=0.91$. Based on the fit of the tilted field data shown in Fig.\,\ref{Fig4_Finite_thickness} to Eq.\,(\ref{eq:Btr_two3rds_full}), one obtains $B_\mathrm{tr}^0/(1+\delta)^2\approx26$\,T. Using $B_\mathrm{tr}^0=(g_e\xi)^{-2}$ (see Eq.\,(\ref{eq:Btr_finite_thickness}) and the related discussion), we get the effective mass parameter, $\xi\equiv\frac{m_\mathrm{CF}}{m_e}\frac{1}{\sqrt{B}}\approx0.4$\,Tesla$^{-1/2}$ (note that $m_\mathrm{CF}=\xi\sqrt{B}m_e$ with $B$ in units of Tesla), which is close to the predicted value ($\xi$=0.6\,Tesla$^{-1/2}$) for the polarization mass, but about 5 times larger than the activation mass~\cite{Park98}. The latter is expected to be relevant in the thermally activated transport measurement in the incompressible regime. In contrast, the transport measurements in this work were carried out in the spin phase transition region where the energy gap is reduced and  the displacement of the phase transition peak is used to obtain $m_\mathrm{CF}$ instead. As a result, the relevant composite fermion mass is the polarization mass, which was measured previously with optical experiments~\cite{Kukushkin99} and NMR~\cite{Dementyev99,Melinte00}.

\section{\label{sec:summary}Summary}
It has been demonstrated in this work that the electric current applied to a 2D electron system at filling factor 1/2 can cause a large change in the degree of nuclear spin polarization. Much of the effect can be attributed to current induced electron heating. This can be described well by the hydrodynamic model. For the current densities applied in this work, the nuclear spin polarization follows a Curie law dependence on the electron temperature.
 The electron heating induced nuclear depolarization effect is mediated by the efficient coupling between the nuclear spin and the electron spin system.

Some advances in nuclear magnetometry have also been made in this work. The finite thickness effect has been included in the study of the spin transition at $\nu$=2/3.  The finite thickness correction is found to be indispensable even for the 16\,nm thick quantum well sample. An alternative nuclear magnetometry technique based on the spin transition at $\nu=1/2$ has also been developed in this work. The results obtained from these two different methods of nuclear magnetometry are consistent with each other.

The capability of manipulating nuclear spin polarization with current as well as the detection of the change in the degree of nuclear spin polarization by electron transport provide a complete toolbox for \textit{all-electrical} nuclear spin relaxation measurements. The advantage of this approach is that the measurement can be carried out under very weak external excitations (a small quasi-dc perturbation without the need for high frequency radiation) and hence at the lowest possible electron temperature. This is highly desirable for studying the spin properties of for instance fragile fractional quantum Hall states.

\begin{acknowledgments}
We are grateful for valuable discussions with N.~R. Cooper, M.~I. D'yakonov and J.~P. Eisenstein. Y.Q.L. acknowledges technical assistance of T. Guan in recording the data for Fig.\,\ref{Fig2_Two3rd_Transition}. We thank B. Frie{\ss}  for valuable comments on the manuscript. We acknowledge financial support from National Science Foundation of China (Project numbers: 10974240 and 91121003), National Basic Research Program of China (Project numbers: 2009CB929101 and 2012CB921703), the Hundred Talents Program of the Chinese Academy of Sciences, the German Ministry of Science and Education and German Israeli Foundation.
\end{acknowledgments}


\end{document}